IEEE*Access*

Multidisciplinary ¦ Rapid Review ¦ Open Access Journal

Date of publication xxxx 00, 0000, date of current version xxxx 00, 0000.

*Digital Object Identifier 10.1109/ACCESS.2020.Doi Number*

# Reliable Tuberculosis Detection using Chest X-ray with Deep Learning, Segmentation and Visualization

**Tawsifur Rahman[1], Amith Khandakar[2], Muhammad Abdul Kadir[1], Khandaker R. Islam[3], Khandaker F. Islam[2], Rashid Mazhar[4,5], Tahir Hamid[6,5], Mohammad T. Islam[7], Zaid B. Mahbub[8], Mohamed Arselene Ayari[9]\*, Muhammad E. H. Chowdhury[2]**

[1]Department of Biomedical Physics & Technology, University of Dhaka, Dhaka-1000, Bangladesh
[2]Department of Electrical Engineering, Qatar University, Doha-2713, Qatar
[3]Department of Orthodontics, Bangabandhu Sheikh Mujib Medical University, Dhaka-1000, Bangladesh
[4]Thoracic Surgery, Hamad General Hospital, Doha-3050, Qatar
[5]Department of Medicine, Weill Cornell Medicine – Qatar, Doha-24811, Qatar
[6]Cardiology, Hamad General Hospital, Doha-3050, Qatar
[7]Department of Electrical, Electronic & Systems Engineering, Universiti Kebangsaan Malaysia, Bangi, Selangor 43600, Malaysia
[8]Department of Mathematics and Physics, North South University, Dhaka-1229, Bangladesh
[9]College of Engineering, Qatar University, Doha-2713, Qatar

\*Correspondence: Mohamed Arselene Ayari; arslana@qu.edu.qa; Tel.: +974 4403 4117

**ABSTRACT:** Tuberculosis (TB) is a chronic lung disease that occurs due to bacterial infection and is one of the top 10 leading causes of death. Accurate and early detection of TB is very important, otherwise, it could be life-threatening. In this work, we have detected TB reliably from the chest X-ray images using image pre-processing, data augmentation, image segmentation, and deep-learning classification techniques. Several public databases were used to create a database of 700 TB infected and 3500 normal chest X-ray images for this study. Nine different deep CNNs (ResNet18, ResNet50, ResNet101, ChexNet, InceptionV3, Vgg19, DenseNet201, SqueezeNet, and MobileNet), which were used for transfer learning from their pre-trained initial weights and trained, validated and tested for classifying TB and non-TB normal cases. Three different experiments were carried out in this work: segmentation of X-ray images using two different U-net models, classification using X-ray images, and segmented lung images. The accuracy, precision, sensitivity, F1-score, specificity in the detection of tuberculosis using X-ray images were 97.07 %, 97.34 %, 97.07 %, 97.14 % and 97.36 % respectively. However, segmented lungs for the classification outperformed than whole X-ray image-based classification and accuracy, precision, sensitivity, F1-score, specificity were 99.9 %, 99.91 %, 99.9 %, 99.9 %, and 99.52 % respectively. The paper also used a visualization technique to confirm that CNN learns dominantly from the segmented lung regions results in higher detection accuracy. The proposed method with state-of-the-art performance can be useful in the computer-aided faster diagnosis of tuberculosis.

**INDEX TERMS**: Tuberculosis detection, TB screening, Deep learning, Transfer learning, Lungs Segmentation, Image processing.

## I. INTRODUCTION

Tuberculosis (TB) is a communicable disease caused by a bacterium called Mycobacterium tuberculosis. It is the leading cause of death from a single infectious disease [1]. Fortunately, this bacterial infectious disease can be well treated by antimicrobial drugs. Early diagnosis of tuberculosis and consequent administration of proper medication can cure this deadly disease [2]. Chest X-rays (CXR) are commonly used for detection and screening of pulmonary tuberculosis [3, 4]. In clinical practice, chest radiographs are examined by experienced physicians for the detection of TB. However, this is time consuming and a subjective process. Subjective inconsistencies in disease diagnosis from radiograph is inevitable [5, 6]. Importantly, CXR images of tuberculosis are often misclassified to other diseases of similar radiologic patterns [7, 8], which may lead to wrong medication to the patients and thereby worsening the health condition. There is also a lack of trained radiologists in the low resource countries (LRC), especially in the rural areas. In this perspective, computer aided diagnosis (CAD) systems can play important role in the mass screening of pulmonary TB by analyzing the chest X-ray images. Recently, artificial intelligence (AI) based solutions have been proposed for many biomedical applications including brain tumor, lungs nodule, pneumonia, breast cancer detection, physiological monitoring and social sensing [9-15] . Among the deep machine learning (ML) techniques, convolutional neural networks (CNNs) have shown great promise in image classification and therefore widely adopted by the research community [16-18]. X-ray radiography is a low-cost imaging technique and there is an abundance of data available for training different machine learning models making deep learning techniques popular for the automatic detection of lung diseases from chest radiographs.

CNNs have been used in several recent studies for the detection of lungs diseases including pneumonia and tuberculosis by analysing chest X-ray images. In response to the COVID-19 pandemic situation in 2020, CNN based techniques have been used for the detection of the novel coronavirus infection form CXR images. Tahir et. al. [19] classified different coronavirus families (SARS, MERS and COVID-19) using transfer learning of various pre-trained CNN models with sensitivity values greater than 90%. Chowdhury et. al. [20] developed a trained model using chest X-ray dataset to distinguish COVID-19 pneumonia, viral pneumonia and normal cases. Chhikara et. al. [17] explored the possibility of detecting pneumonia from CXR images and evaluated the performance of some pre-trained models (Resnet, ImageNet, Xception, and Inception) applying preprocessing techniques like filtering and gamma correction. Abbas et. al [21] reported a modified transfer learning CNN termed as Decompose, Tranfer and Compose (DeTraC) that can deal with data imbalance in medical image classification. This CNN architecture was shown to have improved performance in detecting normal and abnormal x-rays with an accuracy of 99.8 %.

Several research groups used classical machine learning techniques for classifying TB and non-TB cases from CXR images [22-27]. The use of deep machine learning algorithms have been reported in the detection of tuberculosis by varying the parameters of deep-layered CNNs [28-33]. Concept of transfer learning in deep learning framework was used for the detection of tuberculosis utilizing pre-trained models and their ensembles [34-38]. Hooda et. al [28] presented a deep learning approach to classify CXR images into TB and non-TB categories with an accuracy of 82.09%. Evalgelista et al. [30] reported a computer-aided approach based on intelligent pattern recognition using CNNs for TB detection from chest X-ray images with an accuracy of 88.76%. Pasa et al. [31] proposed a deep network architecture optimized for the screening of tuberculosis with an accuracy of 86.82%. They also reported a tool for interactive visualization of TB cases. Nguyen et al. [32] evaluated the performance of a pre-trained model, DenseNet, to classify normal and tuberculosis images from Shenzhen (CHN) and Montgomery County (MC) databases [39] using fine-tuned model, and reported the Area Under the Curve (AUC) values of 0.94 and 0.82 respectively. Hernandez et al. [33] proposed a method for the automatic classification of TB from X-Ray images using an ensemble of CNN models with an accuracy of 86%. Lopes et al. [35] used different pre-trained CNN models to classify the chest radiographs into TB positive and TB negative classes. The performance of the system was evaluated on two publicly available chest X-ray datasets (CHN and MC) and achieved an accuracy of 80%. Meraj et al. [36] used four different CNN models (VGG-16, VGG-19, RestNet50, and GoogLeNet) and explored the limits of accuracies for small-scale and large-scale CNN models in the classification of TB from chest X-rays. Ahsan et al. [37] proposed a generalized pre-trained CNN model for TB detection and achieved accuracies of 81.25% and 80% with and without the application of image augmentation respectively. Yadav et al. [38] reported the detection of tuberculosis using transfer learning technique, which showed an accuracy of 94.89%. Abbas et al. [21] proposed a CNN architecture based on a class decomposition approach to improve the performance of ImageNet pre-trained CNN models using transfer learning and achieved high accuracy in TB detection on Japanese Society of Radiological Technology (JSRT) database. It is worth mentioning that transfer learning methods have also been used to classify the images of TB culture test. Chang et. al. [40] used transfer learning technique on labelled tuberculosis culture images and achieved precision and sensitivity values of 99% and 98% respectively. However, classification of TB culture image requires specific samples from the patients and is not as robust compared to classification from chest X-rays which are readily available.

In clinical applications, an increase in accuracy of TB detection from chest radiographs with a robust and versatile method can make computer aided automatic diagnostic tools more reliable. The classification accuracy can be improved either by using different deep learning algorithms or by modifying the existing outperforming algorithms or combining several outperforming algorithms as an ensemble model. Typically, whole X-ray images were used for the detection of lung disorders using CNN. However, the X-ray images contains lungs as well as other regions of the thorax although the disease like TB is manifested in the lung region only. Thus, focusing on the lung region of the X-ray images during training and classification may significantly improve the performance of TB detection. To the best of our knowledge, no such work regarding the use of deep learning networks on segmented lungs for TB detection is reported. This paper focuses on the detection of TB using transfer learning based technique of CNNs on the original and segmented lungs in X-ray images. CNN based visualization techniques are also implemented to confirm that the deep networks perform classification using the main region of interest, i.e., lungs minimizing learning from irrelevant regions in chest X-rays.

Several important contributions were reported in this study. Firstly, two different U-net models were investigated for the segmentation of the chest X-ray images. Secondly, nine different pre-trained CNNs were applied for the detection of TB from the original and segmented lungs of X-ray images and their performances were analysed. Then, the performance of TB detection by the pre-trained networks using non-segmented X-ray images and segmented images were compared. Finally, score class activation mapping (Score-CAM) based visualization technique was used to demonstrate the regions of the X-ray images that contribute in the classification by the CNNs to confirm whether the segmented lungs X-ray images based classification is more reliable than that of whole X-ray images or not.

The rest of the paper is divided in the following sections: Section 2 summarizes different pre-trained networks for image classification, U-net models for lung segmentation and Score-CAM based visualization techniques. Section 3 describes dataset, pre-processing steps and methodology of this study, while Section 4 summarizes the results of the classification using whole X-ray images and segmented lung images and results are discussed and compared with some other recent studies. Finally, Section 5 concluded the study.

## II. BACKGROUND
### A. DEEP CONVOLUTIONAL NEURAL NETWORKS (CNNS) BASED TRANSFER LEARNING

As discussed earlier, deep CNNs have been popular due to their improved performance in image classification. The convolutional layers in the network along with filters help in extracting the spatial and temporal features in an image. Transfer learning can be useful in those applications of CNN where the dataset is not large. Recently, transfer learning has been successfully used in various field applications such as manufacturing, medical and baggage screening [41-43]. This removes the requirement of having large dataset and also reduces the long training period as is required by the deep learning algorithm when developed from scratch [44, 45].

Nine popular pre-trained deep learning CNNs such as ResNet18, ResNet50, ResNet101 [46], DenseNet201 [47], ChexNet [48], SqueezeNet [49], InceptionV3 [50], VGG19 [51] and MobileNetV2 [52] were used for TB detection. All of these networks apart from ChexNet were initially trained on ImageNet database. Residual Network (in short ResNet) was originally developed to solve vanishing gradient and degradation problem [46]. ResNet has several different variants: ResNet18, ResNet50, ResNet101 and ResNet152 based on the number of layers in the residual network. ResNet was successfully used in biomedical image classification for transfer learning. Typically, deep neural network layers learn low or high level features during training while ResNet learns residuals instead of features [53].

Dense Convolutional Network (in brief DenseNet) needs less number of parameters than a conventional CNN, as it does not learn redundant feature maps. The layers in DenseNet are very narrow, which add a lesser set of new feature-maps. DenseNet has four different common variants: DenseNet121, DenseNet169, DenseNet201 and DenseNet264. Each layer in DenseNet has direct access to the original input image and gradients from the loss function. Therefore, the computational cost significantly reduced, which makes DenseNet a better choice for image classification. ChexNet Pretrained model is a modified version of DenseNet121 and this network is specially trained on large number of chest X-ray images [48].

SqueezeNet and MobileNetv2 are very compact network compared to the other networks. The foundation of SqueezeNet network is a fire module, which consists of Squeeze Layer and Expand layer. The Squeeze layer has only $1 \times 1$ filters, which is feeding to an Expand layer, which has a mixture of $1 \times 1$ and $3 \times 3$ convolution filters [49]. VGG [51] addresses a very important aspect of CNNs, which is depth. The convolutional layers in VGG network use a very small receptive field. There are $1 \times 1$ convolution filters which act as a linear transformation of the input, which is followed by a rectified linear unit (ReLU) layer. The convolution stride is fixed to 1 pixel so that the spatial resolution is preserved after convolution. VGG has different variants: VGG16, and VGG19.

MobileNet structure is built on depth-wise separable convolutions except for the first layer which is a full convolution. All layers are followed by a batch normalization and ReLU nonlinearity with the exception of the final fully connected layer which has no nonlinearity and feeds into a Softmax layer for classification. A final average pooling reduces the spatial resolution to 1 before the fully connected layer. Counting depth-wise and pointwise convolutions as separate layers, MobileNet has 28 layers. Inception modules are used in Convolutional Neural Networks to allow for more efficient computation and deeper networks through a dimensionality reduction with stacked $1 \times 1$ convolutions. The modules were designed to solve the problem of computational expense, as well as over-fitting, among other issues.

### B. SEGMENTATION
There are several variants of segmentation models based on U-nets are available in the literature. Two different variants of called original U-Net [54] and Modified U-Net [55] were investigated in this work to utilize best performing one. Figure 1 shows the architecture of original U-Net and Modified U-Net. The original U-net consists of a contracting path and an expanding path. The contracting path consists of the repeated application of two 3x3 convolutions (unpadded convolutions), each followed by a ReLU and a 2x2 max pooling operation with stride 2 for down sampling. The expanding path consists of an up sampling of the feature map followed by a 2x2 convolution ("up-convolution") that halves the number of feature channels, a concatenation with the correspondingly cropped feature map from the contracting path, and two 3x3 convolutions, each followed by a ReLU. Total 23 convolutional layers are used in the network.

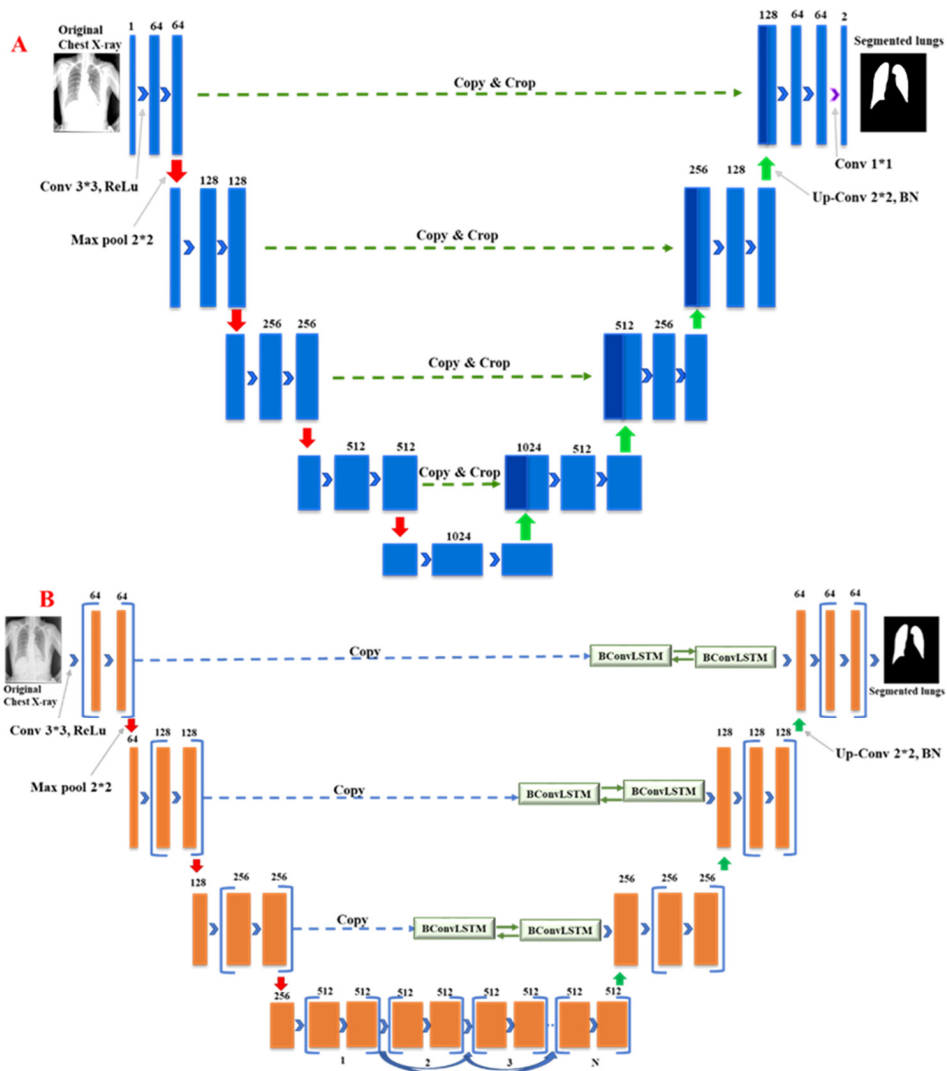

**Figure 1:** Architecture of A) original U-Net and B) modified U-Net.

Modified U-Net also consists of a contracting path and an expanding path as U-Net. The contracting path includes four steps. Each step consists of two convolutional 3×3 filters followed by a 2×2 max pooling function and ReLU. The U-Net might learn redundant features in successive convolution layers. However, modified U-Net uses densely connective convolutions to mitigate this problem. Each step in the expanding path starts by performing an up-sampling function over the output of the previous layer. In the modified U-Net, the corresponding feature maps in the contracting path are cropped and copied to the expanding path. These feature maps are then concatenated with the output of the up-sampling function. Instead of a simple concatenation in the skip connection of U-Net, Modified Unet employs bidirectional Convolutional Long Short Term Memory (BConvLSTM) to combine the feature maps extracted from the corresponding contracting path and the previous expanding up-convolutional layer in a non-linear way.

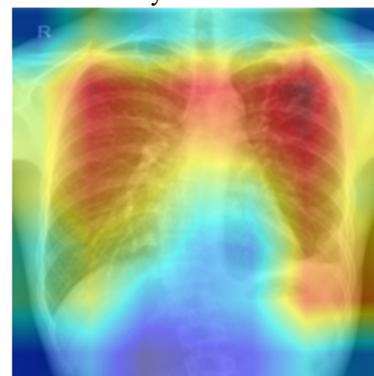

**Figure 2.** Score-CAM heat map on chest X-ray images, where it was shown that different regions of X-ray images were used in decision making by the CNN.

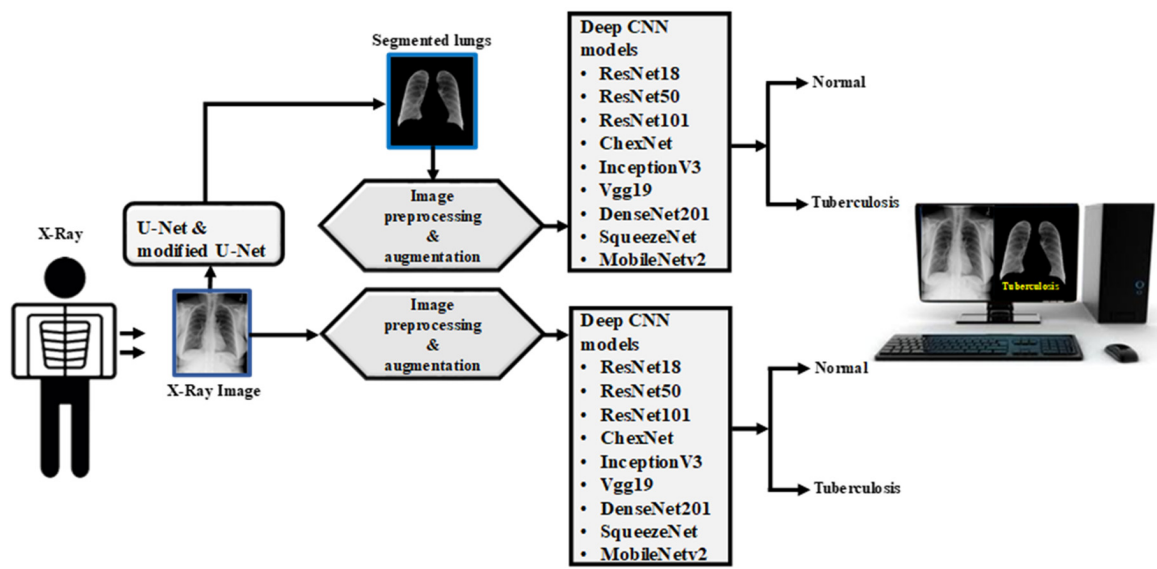

**Figure 3:** Overview of the complete system.

## C. VISUALIZATION TECHNIQUES

Increasing interest on the internal mechanisms of the CNNs and the reasoning behind a network making specific decisions have led to the developments of visualization techniques. The visualization techniques help in better visual representation for interpreting the decision-making process of CNNs. These also increase the model's transparency by visualizing the logic behind the inference that can be interpreted in a way easily understandable to human, thereby increasing the confidence on the outcomes of the neural networks. Amongst the various visualization techniques like SmoothGrad [56], Grad-CAM [57], Grad-CAM++ [58], Score-CAM [59], recently proposed Score-CAM was used in this work due to its promising performance. Score-CAM gets rid of the dependence on gradients by obtaining the weight of each activation map through its forward passing score on target class, the final result is obtained by a linear combination of weights and activation maps. A sample image visualization with Score-CAM is shown in Figure 2, where the heat map indicates that the lungs regions dominantly contributed in the decision making in CNN. This can be helpful to understand how the network is taking its decision and also to improve the confidence of the end-user when it can be confirmed that all the time the network is taking decisions using the lungs of the chest X-rays.

## III. METHODOLOGY

The overall methodology of this study is illustrated using Figure 3. Two different databases were created for this study. One was for lung segmentation while the other one was for TB classification. Three major experiments were carried out in this study. Firstly, two different U-Net models were investigated to identify the suitable network for segmenting lung regions of the X-ray images. Secondly, original chest X-ray images were used for TB classification using nine different pre-trained networks and evaluate classification reliability using Score-CAM technique. Thirdly, segmented lungs of X-ray images were used for TB classification using same networks and evaluated their performance using Score-CAM technique.

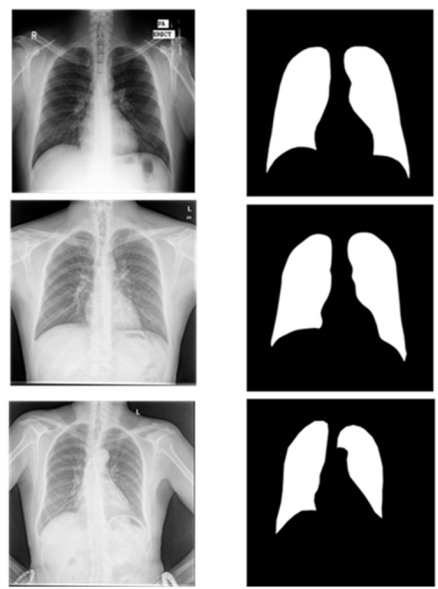

**Figure 4:** Example X-ray images and corresponding ground truth lung masks from Kaggle dataset.

### A. DATASETS DESCRIPTION
#### Lung Segmentation

In this work, Kaggle Chest X-ray images and corresponding lung mask dataset [60] was used for training the lung segmentation models, where 704 X-ray images and their corresponding ground truth lung masks are available. All masks were annotated by expert radiologist; sample X-ray images and masks are shown in Figure 4. There are 360 normal X-ray images and 344 abnormal (infected lung) X-ray images were available in the dataset. Therefore, U-Net networks were trained with both normal and abnormal images.

*TB Classification*

Three publicly accessible databases were used for TB classification problem. These are NLM dataset, Belarus dataset and RSNA dataset:

**NLM dataset:** National Library of Medicine (NLM) in U.S. [61] has made two lung X-ray datasets publicly available: the Montgomery and Shenzhen datasets. The Montgomery County (MC) and the Shenzhen, China (CHN) databases are comprised of 138 and 667 posterior-anterior (PA) chest X-ray images respectively. The resolution of the images of MC database was either 4,020×4,892 or 4,892×4,020 pixels whereas that for CHN database was variable but around 3000×3000 pixels. In the MC database, out of 138 chest X-ray images, 58 images were taken from different TB patients and 80 images were from normal subjects. In the CHN database, out of 662 chest X-ray images, 336 images were taken from different TB patients and 324 images were from normal subjects. Therefore, in this NLM database, there are 406 normal and 394 TB infected X-ray images.

**Belarus dataset:** Belarus Set [62] was collected for a drug resistance study initiated by the National Institute of Allergy and Infectious Diseases, Ministry of Health, Republic of Belarus. The dataset contains 306 CXR images of 169 patients. Chest radiographs were taken using the Kodak Point-of-Care 260 system and the resolution of the images was 2248×2248 pixels. All the images of this database are TB infected images.

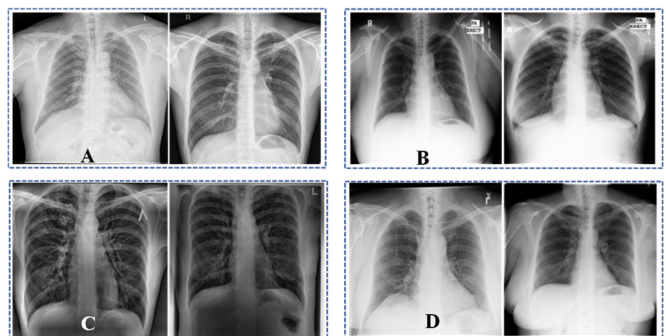

**Figure 5:** Example CXR images from different datasets. (A) CHN dataset, (B) MC dataset, (C) Belarus dataset and (D) RSNA dataset.

**RSNA CXR dataset:** RSNA pneumonia detection challenge dataset [63], which is comprised of about 30,000 chest X-ray images, where 10,000 images are normal and others are abnormal and lung opacity images. All images are in DICOM format. To create a normal database of 3,500 chest X-ray images for this study, 3,094 normal images were taken from this database and rest of the 406 normal images were from NLM database. However, the number of TB infected images by combining NLM and Belarus dataset is 700. Samples of the X-ray images used in the study is shown in figure 5.

**B. PREPROCESSING AND DATA AUGMENTATION**

The size of the input images for different CNNs were different and therefore the datasets were preprocessed to resize the X-Ray images. In segmentation problem, for original U-Net and modified U-Net, the images were resized to 256×256 pixels. In classification problem, for InceptionV3 the images were resized to 227×227 pixels whereas for ResNet, DenseNet, ChexNet, VGG, MobileNetV2 and SqueezeNet, the images were resized to 224×224 pixels. All images were normalized using Z-score normalization using image database mean and standard deviation.

*Data augmentation*

It is reported that the data augmentation can improve the classification accuracy of the deep learning algorithms by augmenting the existing data rather than collecting new data [64]. Data augmentation can significantly increase the diversity of data available for training models. Image augmentation is crucial when the dataset is imbalance. In this study, the number of normal images was 3,500 which is 5 times larger than TB infected images. Therefore, it was important to augment TB infected images four-times to make the database balance. Some of the deep learning frameworks have data augmentation facility built-in with the algorithms, however, in this study, two different image augmentation techniques (rotation, and translation) were utilized to generate TB training images, as shown in Figure 6.

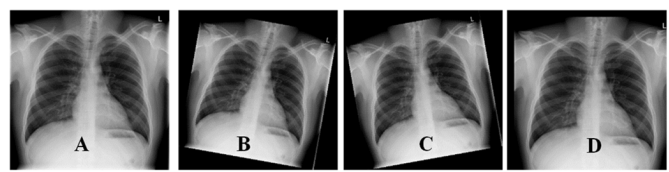

**Figure 6:** Chest X-ray images: original (A), after clockwise rotation by 10 degree (B), after anti-clockwise rotation by 10 degree (C) and after 10% translation (D).

The rotation operation used for image augmentation can be done by rotating the images in the clockwise and counter clockwise direction. Image translation can be done by translating the image in either horizontal (width shift) or vertical direction (height shift) or in both directions. In this work, original image was clockwise and counter clockwise rotated with an angle of 5 and 10 degrees and horizontally and vertically translated by 10% and 15%.

TABLE 1
DETAILS OF TRAINING, VALIDATION AND TEST SET FOR U-NET SEGMENTATION MODELS

| Dataset | No. of CXR images & masks | Train set/fold | Validation set/fold | Test set/fold |
|---|---|---|---|---|
| Kaggle lung x-ray & masks dataset | 704 | 451 | 112 | 141 |

TABLE 2
DETAILS OF TRAINING, VALIDATION AND TEST SET FOR CLASSIFICATION PROBLEM

| Database | Types | Total No. of X-ray images/ class | Training without & with image augmentation | | | |
|---|---|---|---|---|---|---|
| | | | Training set/fold | Augmented train image/fold | Validation /fold | Test image/ fold |
| NLM, Belarus and RSNA | Normal | 3500 | 2240 | 2240 | 560 | 700 |
| | Tuberculosis | 700 | 448 | 2240 | 112 | 140 |

## C. EXPERIMENTS

### Lung Segmentation

Original U-Net and modified U-Net were used separately on Kaggle CXR images and lung mask dataset for lung segmentation. Out of 704 CXR images and lung masks, 80% images and masks were used for training and 20% for testing as summarized in Table 1. Five-fold cross validation was used for training, validation and testing the entire dataset.

The networks were implemented using PyTorch library with Python 3.7 on Intel® Xeon® CPU E5-2697 v4 @ 2,30GHz and 64 GB RAM, with a 16 GB NVIDIA GeForce GTX 1080 GPU. Both the U-Net models were trained using Stochastic Gradient Descent (SGD) with momentum optimizer with learning rate, $\alpha = 10\text{-}3$, dropout rate=0.2, momentum update, $\beta = 0.9$, mini-batch size of 32 images with 50 back propagation epochs; early stopping criterion of 5 maximum epochs when no improvement in validation loss is seen.

### TB Classification

Nine different CNN models were trained, validated and tested separately using non-segmented and segmented X-ray images for the classification of TB and non-TB normal images to investigate if data segmentation can improve classification accuracy. The complete image set were divided into 80% training and 20% testing sub-sets for five-fold cross validation and 20% of training data were used for validation. For example, 80% (2800) of 3500 normal X-ray images were used for training and 20% (700) images were used for testing. However, 20% (560) of 2800 training images were used for validation and therefore, remaining 2240 images were used for training a fold. Table 2 shows the number of training, validation and test images used in the two experiments of non-segmented and segmented lungs images.

All nine CNNs were implemented using PyTorch library with Python 3.7 on Intel® Xeon® CPU E5-2697 v4 @

2,30GHz and 64 GB RAM, with a 16 GB NVIDIA GeForce GTX 1080 GPU. Three comparatively shallow networks (MobileNetv2, SqueezeNet and ResNet18) and six deep networks (Inceptionv3, ResNet 50, ResNet101, CheXNet, VGG19 and DenseNet201) were evaluated in this study to investigate whether shallow or deep networks are suitable for this application. Three different variants of ResNet were used to compare specifically the impact of shallow and deep networks with similar structure. Performance difference due to initially trained on different image classes other than X-ray images were compared with CheXNet, which is a 121-layer DenseNet variant and the only network pre-trained on X-ray images. Nine pre-trained CNN models were trained using same training parameters and stopping criteria as mentioned earlier. However, only 15 back propagation epochs were used for classification problem. Five-fold cross-validation results were averaged to produce the final receiver operating characteristic (ROC) curve, confusion matrix, and evaluation matrices. Using image augmentation and having a validation image set, helps in avoiding overfitting of the models [65].

## D. PERFORMANCE MATRIX

### Lung Segmentation

The performance of different networks in image segmentation for the testing dataset was evaluated after the completion of training and validation phase and was compared using four performance metrics: loss, accuracy, IoU, Dice. The equations used to calculate accuracy, Intersection-Over-Union (IoU) or Jaccard Index and Dice coefficient (or F1-score) are shown in equation (1-3).

$$Accuracy = \frac{(TP+TN)}{(TP+FN)+(FP+TN)} \quad (1)$$

$$IoU(Jaccard\ index) = \frac{(TP)}{(TP+FN+FP)} \quad (2)$$

$$Dice\ Coefficient\ (F1-score) = \frac{(2*TP)}{(2*TP+FN+FP)} \quad (3)$$

*TB Classification*

The performance of different CNNs for testing dataset was evaluated after the completion of training and validation phase and was compared using six performance metrics: accuracy, sensitivity or recall, specificity, precision, area under curve (AUC), F1 score. The matrices were calculated using the following equations (4-8):

$$Accuracy = \frac{(TP+TN)}{(TP+FN)+(FP+TN)} \qquad (4)$$

$$Sensitivity = \frac{(TP)}{(TP+FN)} \qquad (5)$$

$$Specificity = \frac{(TN)}{(FP+TN)} \qquad (6)$$

$$Precision = \frac{(TP)}{(TP+FP)} \qquad (7)$$

$$F1\ Score = \frac{(2*TP)}{(2*TP+FN+FP)} \qquad (8)$$

Here, true positive (TP), true negative (TN), false positive (FP) and false negative (FN) were used to denote number of tuberculosis images identified as tuberculosis, number of normal images identified as normal, number of normal images incorrectly identified as tuberculosis images and number of tuberculosis images incorrectly identified as normal, respectively.

## IV. RESULTS AND DISCUSSIONS
### A. LUNG SEGMENTATION
Original U-Net and modified U-Net networks were trained, validated and evaluated on test data for the segmentation of lungs of X-ray images. Table 3 shows comparative performance of both U-Net models in image segmentation. Figure 7 shows some example test X-ray images, corresponding ground truth masks and segmented lung images generated by the two trained U-Net models for kaggle dataset. It can be noted that the original U-Net outperformed modified U-Net in the segmentation of lung regions on CXR images quantitatively and qualitatively.

TABLE 3
COMPARATIVE PERFORMANCE OF ORIGINAL U-NET AND MODIFIED U-NET

| Network | Test loss | Test accuracy | IoU | Dice |
|---|---|---|---|---|
| U-Net | 0.039 | 98.03 | 92.4 | 96.0 |
| Modified U-Net | 0.043 | 97.88 | 91.8 | 95.6 |

The better performing original U-Net model was then used to segment the classification database (3500 normal and 700 TB images), which was used for classification of TB and non-TB normal cases. It is important to see on a completely unseen image-set with TB infection and normal images how well the trained segmentation model works. It can be seen from Figure 8 that the original U-net model trained on Kaggle chest X-ray dataset can segment the lung areas of the X-ray images of the classification database very reliably. However, there is quantitative evaluation on the classification dataset is not possible as there is no ground truth masks available for this database and therefore, qualitative evaluation were done to confirm that each X-ray image was segmented correctly.

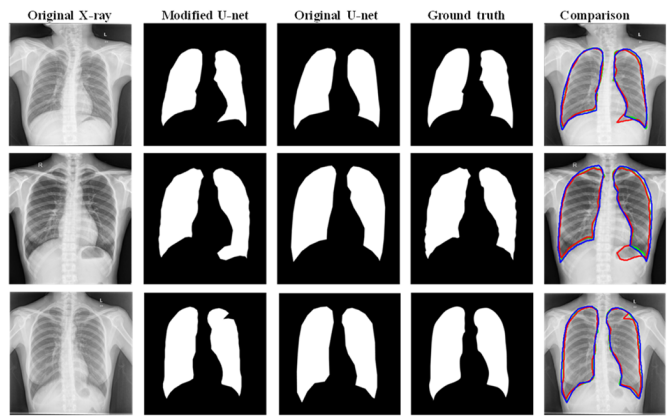

**Figure 7:** Sample test X-ray images, segmented lungs using two U-net models and ground truth were compared.

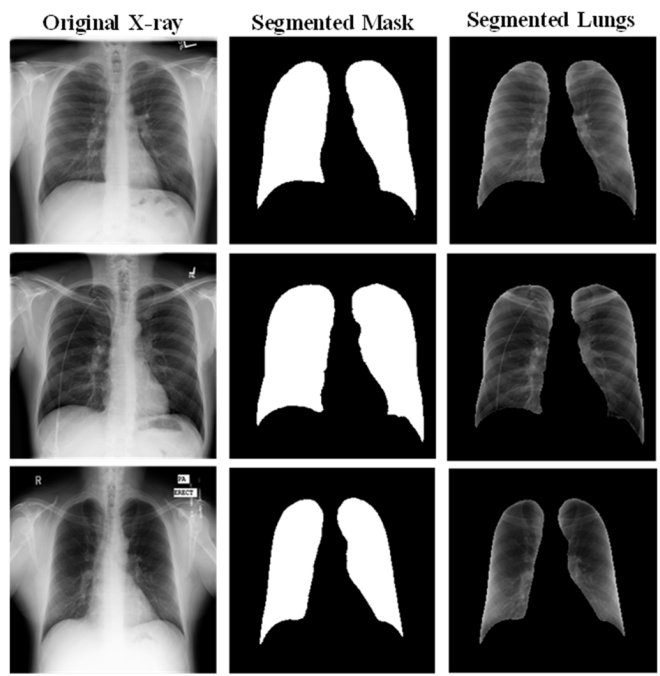

**Figure 8:** Samples X-ray images from classification database (left), generated masks by the trained original U-Net model (middle) and corresponding segmented lung (right).

| Scheme | Model | Overall Accuracy | Weighted Average | | | |
|---|---|---|---|---|---|---|
| | | | Precision | Sensitivity | F1-score | Specificity |
| Without Segmentation (Whole X-ray) | ResNet18 | 95.57 | 96.12 | 95.57 | 95.71 | 95.80 |
| | ResNet50 | 95.86 | 96.39 | 95.86 | 95.99 | 96.54 |
| | **ResNet101** | **96.88** | **97.19** | **96.88** | **96.96** | **97.32** |
| | **ChexNet** | **97.07** | **97.34** | **97.07** | **97.14** | **97.36** |
| | InceptionV3 | 95.95 | 96.40 | 96.00 | 96.20 | 96.89 |
| | Vgg19 | 95.81 | 96.38 | 95.81 | 95.95 | 96.76 |
| | DenseNet201 | 95.83 | 96.47 | 95.84 | 95.99 | 97.46 |
| | SqueezeNet | 95.45 | 96.14 | 95.45 | 95.62 | 96.69 |
| | MobileNet | 95.26 | 95.9 | 95.26 | 95.43 | 95.62 |
| With Segmentation (Segmented lung) | ResNet18 | 99.86 | 99.85 | 99.85 | 99.85 | 99.63 |
| | ResNet50 | 99.88 | 99.88 | 99.88 | 99.88 | 99.52 |
| | ResNet101 | 99.79 | 99.79 | 99.79 | 99.78 | 99.39 |
| | ChexNet | 99.69 | 99.69 | 99.69 | 99.69 | 98.68 |
| | InceptionV3 | 99.83 | 99.83 | 99.83 | 99.83 | 99.62 |
| | Vgg19 | 99.88 | 99.88 | 99.88 | 99.88 | 99.41 |
| | **DenseNet201** | **99.90** | **99.91** | **99.90** | **99.90** | **99.52** |
| | SqueezeNet | 99.67 | 99.67 | 99.66 | 99.66 | 98.56 |
| | MobileNet | 99.76 | 99.76 | 99.76 | 99.76 | 99.15 |

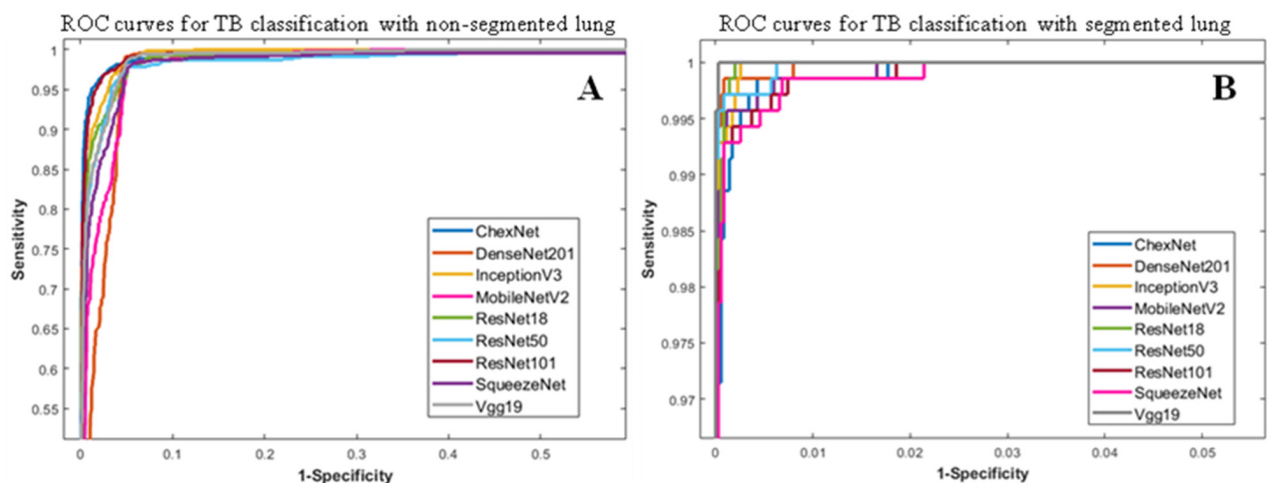

**Figure 9:** Comparison of the ROC curves for Normal, and Tuberculosis classification using CNN based models for non-segmented (A) and segmented (B) CXR images.

## B. TB CLASSIFICATION

As mentioned earlier, there are two different experiments (non-segmented and segmented lungs X-ray images) were conducted for classification of TB and normal (non-TB) cases. The comparative performance for different CNNs for two-class classification between TB and normal images for non-segmented and segmented images is shown in Table 4. It is apparent from Table 4 that all the evaluated pre-trained models perform very well in classifying TB and normal images in two-class problem. Among the networks trained with X-ray images without segmentation, ResNet101 and CheXNet are equally performing better for classifying images. Even though CheXNet is shallower than DenseNet201, it was originally trained on X-ray images provides it additional benefits in classifying X-ray images into two classes and it is showing better performance than DenseNet201. It is not necessary that deeper network will perform better rather CheXNet is a very good example of transfer learning and it outperforms other networks for this problem. Similar performance was observed by the authors in their COVID-19 classification problem [20]. However, ResNet 18, 50 and 101 showed increasingly better performance for the classification of images without segmentation. This clearly reveals that residual network structure provide better performance with network depth which cannot be generalized for VGG19 and InceptionV3 networks. Interestingly, the performance of the shallow networks like SqueezeNet and MobileNetv2 are comparable to most of the deep networks. Figure 9(A) clearly shows that the ROC curves for CheXNet and ResNet101 are overlapping while the ROC curves for rest of the networks are comparable among themselves.

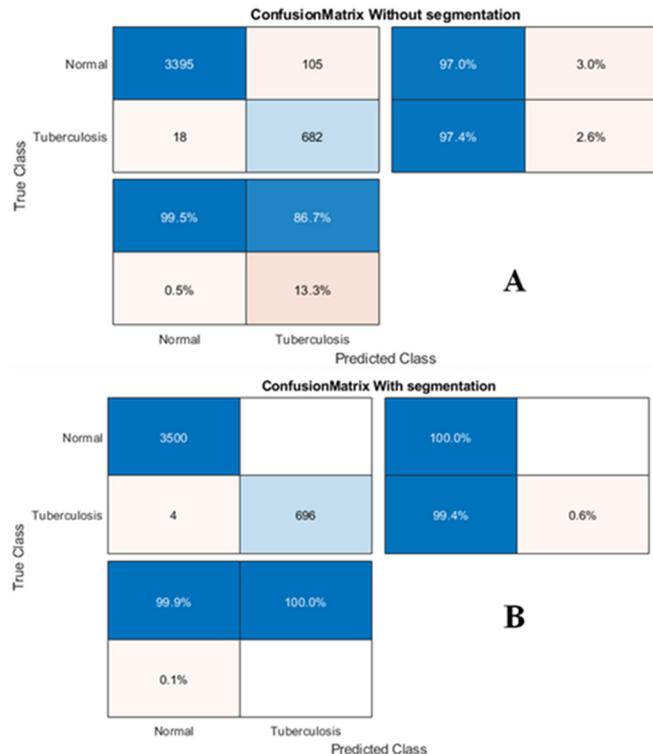

**Figure 10:** Confusion matrix for Normal and Tuberculosis classification for ChexNet model without segmented X-ray (A), and DenseNet201 model with segmented X-ray (B).

On the other hand, lung segmented chest X-ray images provided a clear performance boost for all of the tested networks. This reflect the fact that shallow or deep all CNNs can distinguish TB and normal lungs with very high reliability when only lung images are used as input to the CNNs. Although all the CNNs provide better performance in classification, DenseNet201 showed the most outstanding performance in classifying TB and normal images. In summary, CheXNet and DenseNet201 are producing the highest classification accuracies of 97.07% and 99.9% for non-segmented and segmentation images respectively. DenseNet201 is performing well on the segmented lungs, which reflects that the deeper network can classify more accurately for segmented lung X-ray images. It is evident that segmentation improves overall performance of classification significantly. However, in this classification problem only two class problem (TB and normal images) was considered and the lungs regions of TB images are significantly different than that of normal images and therefore, all the tested networks performed well. This is apparent from the ROC curves of Figure 9 as well. In Figure 9(B), the ROC curves are showing a comparable performance from all the networks.

Figure 10 shows the confusion matrix for outperforming ChexNet model without segmented X-ray images and DenseNet201 model with segmented lungs X-ray images. It can be noticed that even with the best performing network, 18 out of 700 TB images were miss-classified as normal and 105 out of 3500 normal X-ray images were miss-classified as TB image when the non-segmented X-ray images were used as input to the classifier. On the other hand, only 4 out of 700 TB images were miss-classified as normal and no normal images were miss-classified as TB when the segmented lungs were used as input to the classifier. This is clearly an outstanding performance from any computer aided classifier and this can significantly help in the fast diagnosis of TB by the radiologist immediately after acquiring the X-ray images.

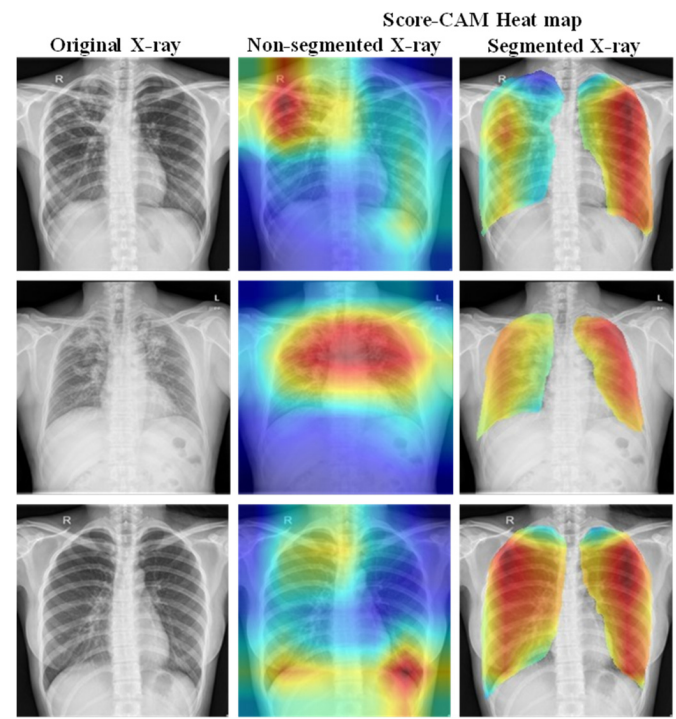

**Figure 11:** Score-CAM visualization of correctly classified TB infected chest X-ray: Original X-ray (left), Score-CAM heat map on original X-ray (middle) and Score-CAM heat map on segmented lungs (right).

### C. VISUALIZATION USING SCORE-CAM

As mentioned earlier, it is important to see where network is learning for the relevant area of the X-ray images or it is learning from anywhere and any non-relevant information for classification. Therefore, Score-CAM based heat maps were generated for original (non-segmented) X-ray images and segmented X-ray images. Figure 11 shows the original X-ray samples along with the heat maps on non-segmented and segmented lung. In each of the non-segmented images, CNN is learning from the regions other than the lungs and the areas which are mostly contributing to take decision are not part of the lungs always or most of the cases. Therefore, even though the performance of the CNNs is quite good in classification of TB and normal images, the reliability of these networks would be criticized and several researchers have recently criticized CNNs for learning from non-relevant areas of the image [66]. However, classification using segmented lungs can overcome this drawback and it is evident from Figure 11 that the heat maps for segmented lungs indicate that the dominantly contributing region in the decision making of CNN are within the lung. As TB changes opacities of the lung regions on the CXR, thus learning by the deep networks with segmented lung images provided higher classification accuracies. Since network is now only learning from lung areas, it can learn differences of normal and TB infected lung images only and therefore, it can distinguish them with very high accuracy.

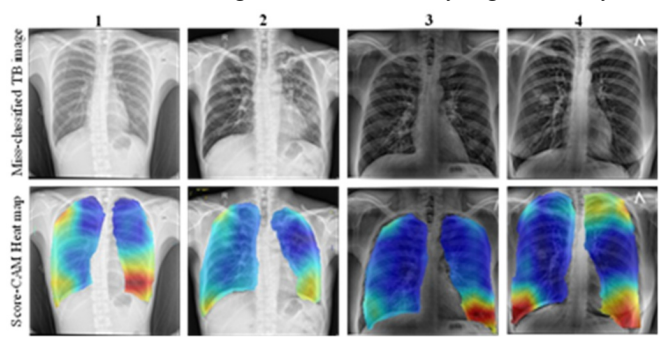

**Figure 12:** Miss-classified TB infected images and their corresponding Score-CAM heat map.

As mentioned earlier, out 3500 normal X-ray images, none was miss-classified to TB infected image while 4 out of 700 TB images were miss-classified to normal by the CNN model trained with segmented lungs. It is therefore, important to investigate these four miss-classified images whether these are early stage of TB and therefore, CNN consider them normal or there is any potential reason of this miss-classification. Figure 12 shows that in all of the four miss-classified images CNN took decision from the lower edge of the lungs and only from a smaller area of lungs and that portion of the lungs was normal for all of the images. First case of TB miss-classified image was very similar to normal image and radiologist was classified them as normal and therefore, it may be very early stage of TB as it was labelled as TB infected image in the original dataset. However, second to fourth TB infected images are from mild to moderate TB patients and CNN miss-classified them as it is learning from wrong area of the lungs. Therefore, if the segmented lungs can be further segmented into patches which can be used as input to CNN model, which might further enhance the performance. This is the future direction of investigation of this work.

This state-of-the-art performance of our proposed method was compared with the recently published works in the same problem domain. Table 5 summarizes the comparison of the results presented in this paper to that of others for the detection of tuberculosis from chest X-ray images. In few studies [67, 68], the detection accuracies have been reported to be 99.8% using databases consisting of small number of images. However, in our study we used larger datasets than others and found consistent results. We also used image segmentation techniques and evaluated classification performance using nine different CNN models that makes our method more robust and versatile with 99.9% accuracy. Moreover, the performance of our model is justified using Score-CAM based visualization technique to emphasize the importance of segmentation of X-ray images for CNN based classification tasks.

TABLE 5
COMPARISON OF FINDINGS OF THIS STUDY WITH OTHER RECENT SIMILAR WORKS

| Author | Year | Method | Database | Evaluation Matrix |
|---|---|---|---|---|
| Hrudya et al. [22] | 2015 | Support Vector Machine (SVM) | MC | Not stated |
| JAIME ET AL. [23] | 2016 | Statistical Analysis | 392 records collected from Cape Town in South Africa | AUC - 0.84, Specificity - 49% and Sensitivity 95 % |
| Rahul et al. [28] | 2017 | CNN | MC and CHN | Accuracy 82.09 % |
| Anuj et al. [34] | 2017 | CNN Transfer Learning | MC and CHN | Accuracy > 80 % |
| Lopes et al. [35] | 2017 | CNN Transfer Learning | MC and CHN | Accuracy 84.7% and AUC - 0.926 |
| Abbas et al. [67] | 2018 | Knowledge transferred via Alexnet | MC | AUC – 99.8 |
| Lucas et al. [30] | 2018 | CNN and two ensembles | JSRT,MC and CHN | Accuracy 88.76 % |
| Ojasvi et al. [38] | 2018 | Transfer learning (ResNet) | MC and CHN | Accuracy 94.89 % |
| Niharika et al. [24] | 2019 | Support Vector Machine | MC and CHN dataset | AUC - 0.96 and specificity - 100% |
| Pasa et al. [31] | 2019 | Optimized CNN | MC, CHN and Belarus Dataset | AUC - 0.811 for MC, 0.9 for CHN and 0.925 for combined |
| Syeda et al. [36] | | Transfer Learning | MC and CHN | AUC-0.85 |
| Mostofa et al. [37] | 2019 | Transfer Learning (VGG 16 model) | MC and CHN | Accuracy 80% and 81.25% without and with augmentation |
| Quang et al. [32] | 2019 | Tuning of DenseNet model | MC and CHN | AUC-0.94 for CHN and 0.82 for MC |
| Alfonso et al. [33] | 2019 | 3 pre-trained CNNs | MC and CHN | Accuracy 86% |
| Abbas et al. [21] | 2020 | DeTraC: Class decomposition | JSRT | Accuracy 99.8% |
| **This paper** | **2020** | **Nine pre-trained CNN models with lungs segmentation.** | **NLM, Belarus & RSNA** | **Accuracy 99.9%** |

## V. CONCLUSION

This work presents a transfer learning approach with deep Convolutional Neural Networks for the automatic detection of tuberculosis from the chest radiographs. The performance of nine different CNN models were evaluated for the classification of TB and normal CXR images. ChexNet model outperforms other deep CNN models for the datasets without image segmentation whereas DenseNet201 outperforms for the segmented lungs. The classification accuracy, precision and recall for the detection of TB were found to be 97.07%, 97.34%, and 97.07% without segmentation and 99.9%, 99.91% and 99.9% with segmentation respectively. It was also shown that image segmentation can significantly improve classification accuracy. The Score-CAM visualization output confirms that lung segmentation helps in taking decisions from the lung region unlike the original x-rays where decision can be taken based on features outside the lung region. Therefore, segmentation of lungs is very crucial for computer aided diagnosis using radiographs. This state-of-the-art performance can be a very useful and fast diagnostic tool, which can save significant number of people who died every year due to delayed or improper diagnosis.

## AUTHORS CONTRIBUTION

Experiments were designed by TR and MEHC. Experiments were performed by TR, AK, KFI and KRI. Results were analyzed by KRI, RM, TH, MEHC, MTI, ZBM and MAK. All the authors were involved in the interpretation of data and paper writing and revision of the article.

## ACKNOWLEDGMENTS

The publication of this article was funded by the Qatar National Library.

## FUNDING

## CONFLICTS OF INTEREST

The authors declare no conflict of interest.